\documentclass[showpacs,aps,prd,nofootinbib,floatfix,amsmath,amssymb]{revtex4}
\usepackage{graphicx}
\begin{document}
\makeatletter
\newbox\slashbox \setbox\slashbox=\hbox{$/$}
\newbox\Slashbox \setbox\Slashbox=\hbox{\large$/$}
\def\pFMslash#1{\setbox\@tempboxa=\hbox{$#1$}
  \@tempdima=0.5\wd\slashbox \advance\@tempdima 0.5\wd\@tempboxa
  \copy\slashbox \kern-\@tempdima \box\@tempboxa}
\def\pFMSlash#1{\setbox\@tempboxa=\hbox{$#1$}
  \@tempdima=0.5\wd\Slashbox \advance\@tempdima 0.5\wd\@tempboxa
  \copy\Slashbox \kern-\@tempdima \box\@tempboxa}
\def\FMslash{\protect\pFMslash}
\def\FMSlash{\protect\pFMSlash}
\def\miss#1{\ifmmode{/\mkern-11mu #1}\else{${/\mkern-11mu #1}$}\fi}
\makeatother

\title{The decay $b\to s\gamma$ in the presence of a constant antisymmetric tensor field}
\author{G. Ahuatzin$^{(a)}$, I. Bautista$^{(b)}$, J. A. Hern\' andez-L\' opez$^{(c)}$, F. Ram\'\i rez-Zavaleta$^{(d)}$, J. J. Toscano$^{(c)}$}
\address{
$^{(a)}$Instituto de F\'\i sica, Universidad Aut\' onoma de San Luis Potos\'\i,
\' Alvaro Obreg\' on 64, Zona Centro, San Luis Potos\'\i, S.L.P. 78000, M\' exico.\\
$^{(b)}$IGFAE and Departamento de F\'\i sica de Part\'\i culas, Univ. of Santiago
de Compostela, 15706, Santiago de Compostela, Spain.\\
$^{(c)}$Facultad de Ciencias F\'{\i}sico Matem\'aticas,
Benem\'erita Universidad Aut\'onoma de Puebla, Apartado Postal
1152, Puebla, Pue., M\'exico.\\
$^{(d)}$Facultad de Ciencias F\'\i sico Matem\' aticas,
Universidad Michoacana de San Nicol\' as de Hidalgo, Avenida
Francisco J. M\' ujica S/N, 58060, Morelia, Michoac\'an, M\'
exico.}

\begin{abstract}
A constant antisymmetric 2-tensor can arise in general relativity with spontaneous symmetry breaking or in field theories formulated in a noncommutative space-time. In this work, the one-loop contribution of a nonstandard $WW\gamma$ vertex on the flavor violating quark transition $q_i\to q_j\gamma$ is studied in the context of the electroweak Yang-Mills sector extended with a Lorentz-violating constant 2-tensor. An exact analytical expression for the on shell case  is presented. It is found that the loop amplitude is gauge independent, electromagnetic gauge invariant, and free of ultraviolet divergences. The dipolar contribution to the $b\to s\gamma$ transition together with the experimental data on the $B\to X_s\gamma$ decay is used to derive the constraint $\Lambda_{LV}>1.96$ TeV on the Lorentz-violating scale.
\end{abstract}

\pacs{13.40.Gp, 14.65.Fy, 11.10.Nx}

\maketitle

\section{Introduction}
\label{I}At present, there is interest for studying possible deviations of the Lorentz symmetry, mainly since Kostelecky and Samuel discovered that certain mechanism in string theory can cause a spontaneous breaking of this symmetry~\cite{KS}. Some mechanisms of violation of Lorentz symmetry have also been found within the context of quantum gravity~\cite{QG}. Since these theories have not been sufficiently developed, an effective field theory that contain both the Standard Model (SM) and gravity has been formulated. This effective theory, for which exists a minimal version without gravity~\cite{SME}, is called the Standard Model Extension (SME)~\cite{SMEG}. Although motivated from specific scenarios in the context of string theory or general relativity with spontaneous symmetry breaking, the SME is beyond these specific ideas due to its generality, which is the main advantage of effective field theories. Thus the SME provides us a powerful tool for investigating Lorentz violation in model-independent manner. Lorentz violation also arises within the context of field theories formulated in a noncommutative space-time~\cite{Snyder}. The idea that the space-time might be noncommutative at short distances or very high energies has been the subject of renewed interest recently due to the Seiberg-Witten's string inspired results~\cite{SWM}, which allow one to connect commutative and noncommutative gauge theories. A method to formulate the noncommutative SM (NCSM) as an effective theory, which is expressed in powers of the noncommutativity parameter, has been proposed in Refs.~\cite{NCYM,NCSM}. The effective theory that results from this formulation of field theories is less wide than the SME, as it arises from the specific idea of noncommutativity of the space-time. Indeed, as it has been shown in Ref.~\cite{SME-NCSM}, the NCSM is a subset of the SME. Although these effective theories introduce constant background fields that carry Lorentz indices, they are not Lorentz invariants under general Lorentz transformations, but only under observer Lorentz transformations. As it has been discussed in reference~\cite{SME-NCSM}, there are two distinct classes of Lorentz transformations, namely, the observer and particle Lorentz transformations. The former class of Lorentz transformations corresponds to a change of coordinates, whereas the latter can be associated with a change of the measurement apparatus~\cite{Rev}.

A violation of Lorentz symmetry would clearly be a dramatic indication of new physics. Since this class of physical phenomenon could be originated at very high energies, their low-energy manifestation only could be detected in those processes that are very suppressed or forbidden within the SM. Precision observables constitute also an unique laboratory to test the standard model (SM). Is the case of the $B\to X_s\gamma$ decay, which is potentially sensitive to new physics effects and has been measured with good accuracy, showing no deviations from the SM~\cite{Review}. This means that this observable may provide stringent constraints on physics beyond the electroweak scale. At the quark level, the $b\to s\gamma$ transition contributes to $B\to X_s\gamma$ via a dipolar electromagnetic transition. In the SM, this radiative process first arises at the one-loop order, but still has a relatively large branching ratio because of the nondecoupling effect of the top quark loop and the large CKM $V_{tb}V^*_{ts}$ factor. The rare $b\to s\gamma$ decay has shown to be very sensitive to possible new physics effects in diverse scenarios~\cite{Hewett}. Its sensitivity to new physics effects has been studied in countless approaches beyond the SM, as supersymmetric models~\cite{SUSY}, the two Higgs doublet model~\cite{2HDM}, left-right symmetric models~\cite{LRM}, technicolor models~\cite{Tech}, models with a fourth generation~\cite{FG}, supergravity models~\cite{SG}, the littlest Higgs model~\cite{LHM}, unparticle interactions~\cite{UP}, $331$ models~\cite{331}, and extra dimensions~\cite{ED}. In particular, the sensitivity of this observable to new physics effects via an anomalous $WW\gamma$ vertex has been investigated by some authors~\cite{Chia,Peterson,Rizzo,McKellar,Toscano} in the context of the effective Lagrangian approach~\cite{EL}. This effective field theory (which from now on will be called conventional effective field theory (CEFT) for comparison purposes) differs from both the SME and the NCSM, as it is formulated under the assumption that it respects simultaneously both the Lorentz and the gauge symmetries. Below, we will see that there are some interesting differences among the SME, the NCSM, and the CEFT.

In this work, we are interested in investigating the sensitivity of the $b\to s\gamma$ transition to an anomalous $WW\gamma$ vertex that can arise, in the context of both the SME and the NCSM, from an electroweak Yang-Mills sector coupled with a constant antisymmetric 2-tensor $b^{\alpha \beta}$. This constant background field that arises in the context of the NCSM as measurement of the noncommutativity of the space-time (usually established as $[x^\alpha,x^\beta]=i\theta^{\alpha \beta}$~\cite{NCSM}) or as a vacuum expectation value of a tensor field $B^{\alpha \beta}$~\cite{SSBQG} in general relativity, has not been still considered in the SME up to now, as only have been introduced observer invariant interactions of dimension lower than four that not involve constant objets with two indices. The fact that this constant background field arises naturally in both of these formulations of Lorentz violation, constitutes an important incentive for studying some of its possible low-energy manifestations. Another motivation for studying Lorentz violation via this tensor arises from the fact that it can couples with a dimension-six $SU_L(2)\times U_Y(1)$-invariant operator constructed only with the strength tensor associated with $SU_L(2)$, which, in its Lorentz invariant version, has been the subject of important interest  in the literature~\cite{Chia,Peterson,Rizzo,McKellar,Toscano, NT}. On the other hand, as discussed in the context of string theory quantization~\cite{SWM} and in general relativity with spontaneous symmetry breaking~\cite{SSBQG}, there exists more than a simple analogy between the six $b^{\alpha \beta}$ quantities and the six components of the electromagnetic field tensor $F^{\alpha \beta}$. For any practical purpose, the dimensionless constant background fields $e^i\equiv \Lambda^2_{LV} b^{0i}$ and $b^i\equiv (1/2)\Lambda^2_{LV}\epsilon^{ijk}b^{jk}$, with $\Lambda_{LV}$ the new physics scale, play the role of an external agent that would induce deviations from the SM predictions, which in principle could be observed in future high-energy experiments.

As mentioned, we will focus on the Yang-Mills part of the effective Lagrangian that characterizes the SME (or also the NCSM) modified by the presence of an observer invariant that arises from the contraction of $b^{\alpha \beta}$ with a Lorentz 2-tensor that is invariant under the $SU_L(2)$ gauge group.  This extended Yang-Mills sector generates a nonrenormalizable $WW\gamma$ vertex, which differs substantially from the one studied in references~\cite{Chia,Peterson,Rizzo,McKellar,Toscano} within the context of the CEFT~\cite{EL}. As we will see below, the anomalous $WW\gamma$ vertex arises in the SME (or in the NCSM) from a dimension-six operator which is invariant under the $SU_L(2)$ gauge group but it is a $2$-tensor under the Lorentz group, whereas in CEFT this operator is invariant under both the gauge group and the Lorentz group. Technically speaking, this means that only the antisymmetric part in the two Lorentz indices contributes to physical amplitudes in the SME due to the antisymmetry of $b^{\alpha \beta}$, whereas only the symmetric part contributes in the CEFT, as in this case the Lorentz invariant operator is obtained through a contraction with the metric tensor $g^{\alpha \beta}$ instead of $b^{\alpha \beta}$. This facts make essentially different our analysis from those presented in references~\cite{Chia,Peterson,Rizzo,McKellar,Toscano}. Although our main purpose in this work is to study the impact of Lorentz violation on the $b\to s\gamma$ decay, we will present exact formulae for the general on-shell $q_i\to q_j\gamma$ process. As we will see below, the corresponding vertex functions are independent of the gauge-fixing procedure used to define the $W$ propagator, electromagnetic gauge invariant, and free of ultraviolet divergences.

The rest of the paper has been organized as follows. In Sec.~\ref{YM}, the extended Yang-Mills sector of the SME or the NCSM at first order in the background tensor field $b^{\alpha \beta}$ is presented and the Feynman rule for the $WW\gamma$ vertex is calculated. The differences that present this Lagrangian with respect the one studied in CEFT are discussed. Sec.~\ref{D} is devoted to calculate the amplitude for the $q_i\to q_j\gamma$ decay. Special attention is put on some features of the amplitude, such as gauge independence and electromagnetic gauge invariance, as well as its finite character. In Sec.~\ref{B} the contribution of the $b\to s\gamma$ transition to the $B\to X_s\gamma$ process is calculated and used to determine a lower bound for the Lorentz violation scale. Finally, in Sec.~\ref{C} the conclusions are presented.

\section{The effective Yang-Mills Lagrangian}
\label{YM}To begin with, we present a brief discussion on the main differences among the SME, the NCS, and the CEFT formulations of effective field theories. As mentioned in the introduction, the extended electroweak Yang-Mills sector that we will consider here can arise in both the SME and the NCSM, the latter being a subset of the former~\cite{SME-NCSM}. Therefore let us first to compare the NCSM with the CEFT. The NCSM is characterized by an effective Lagrangian of the way~\cite{NCSM}\footnote{In this part of our presentation, we will use $\theta^{\alpha \beta}$ instead of $b^{\alpha \beta}$, as it is the notation used in the literature related with the NCSM.}
\begin{equation}
\label{el} {\cal L}_{NCSM}={\cal L}_{SM}+\theta^{\alpha
\beta}\sum^{N_6}_{i=1}{\cal O}^{(6)}_{i \, \alpha \beta}+\theta^{\alpha \beta}\theta^{\mu
\nu}\sum^{N_8}_{i=1}{\cal O}^{(8)}_{i \, \alpha \beta \mu \nu}+\cdots,
\end{equation}
where ${\cal L}_{SM}$ is the usual SM Lagrangian. Here ${\cal O}^{(6)}_{i \, \alpha \beta}$, ${\cal O}^{(8)}_{i \, \alpha \beta \mu \nu}$, $\cdots$, are sets of $N_6$, $N_8$, $\cdots$, $SU_C(3)\times SU_L(2)\times U_Y(1)$-invariant operators of canonical dimension $6$, $8$, $\cdots$, and Lorentz tensors of rank $2$, $4$, $\cdots$, respectively, which couple to the constant background tensor $\theta^{\alpha \beta}$. It should be noticed that due to the antisymmetric character of $\theta^{\alpha \beta}$ only the antisymmetric part of the gauge invariant  Lorentz tensors ${\cal O}^{(6)}_{i \, \alpha \beta}$, ${\cal O}^{(8)}_{i \, \alpha \beta \mu \nu}$, $\cdots$ can contribute. As commented in the introduction, a CEFT is constructed with nonrenormalizable operators ${\cal O}^{(n)}$ of dimension $n>4$ that are invariant under both the gauge and the Lorentz groups:
\begin{equation}
{\cal L}_{CEFT}={\cal L}_{SM}+\sum_{n=5}\sum^{N_n}_{i=1}\frac{\alpha^n_i}{\Lambda^{n-4}}{\cal O}^{(n)}_{i}.
\end{equation}
It is not difficult to convince ourselves that the symmetric part
of the Lorentz tensors appearing in ${\cal L}_{NCSM}$, defined
through the contraction with the metric tensor ${\cal
O}^{(6)}_i=g^{\alpha \beta}{\cal O}^{(6)}_{i \, \alpha \beta}$,
${\cal O}^{(8)}_i=g^{\alpha \beta}g^{\mu \nu}{\cal O}^{(8)}_{i \,
\alpha \beta \mu \nu}+\cdots$, are present in the CEFT, as  ${\cal
L}_{CEFT}$ contains all the operators that respect both the gauge
and Lorentz symmetries~\cite{EL}. However, as already commented,
it is important to stress that the contribution to physical
observables of the operators appearing in ${\cal L}_{CEFT}$ would
differ from those in ${\cal L}_{NCSM}$, as in the former case only
contributes the symmetric part of the  ${\cal O}^{(6)}_{i \,
\alpha \beta}$, ${\cal O}^{(8)}_{i \, \alpha \beta \mu \nu}$,
$\cdots$ Lorentz tensors, whereas in the latter only contributes
the antisymmetric part. This point will be clarified below for the
specific case of the Yang-Mills sector.

We now are in position of pointing out the main differences
between the SME and the CEFT. Technically, the difference between
the SME and the CEFT is that although both theories are made of
gauge invariant operators, they carry Lorentz indices in the
former case but not in the latter. In contrast with the NCSM, all
the gauge invariant operators of the SME that carry an odd number
of Lorentz indices cannot be converted in Lorentz scalars through
contractions with the metric tensor and thus no link with the
elements of the CEFT can be established. On the conceptual side,
the differences between both type of formulations are even more
profound, as the SME represents a model-independent low-energy
formulation of Lorentz violation that incorporates together the SM
and general relativity. On the other hand, the CEFT incorporates
also in a model-independent way virtual effects of physics beyond
the electroweak scale, which respects both the Lorentz symmetry
and the SM gauge symmetry. Thus, by construction, the CEFT
formulation does not incorporate Lorentz violation.

\begin{figure}
\centering\includegraphics[width=1.5in]{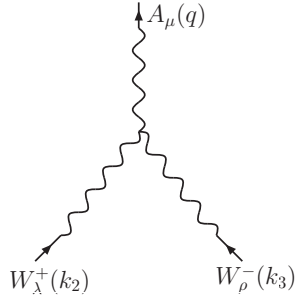}
\caption{\label{V}The trilinear $WW\gamma$ vertex in presence of the background field $b^{\alpha \beta}$.}
\end{figure}

As already mentioned, we are interested in studying the one-loop contribution of the extended Yang-Mills sector to the $q_i\to q_j\gamma$ transition. At first order in $b^{\alpha \beta}$, the extended Lagrangian for the Yang-Mills sector associated with the $SU_L(2)$ group can be written as follows~\cite{NCSM}\footnote{The other possible observer invariant $b^{\alpha \beta}Tr[W_{\alpha \beta}W_{\mu \nu}W^{\mu \nu}]$ vanishes. In fact, $ W^b_{\mu \nu}W^{c\mu \nu}Tr[\sigma^a \sigma^b \sigma^c]=2i W^b_{\mu \nu}W^{c\mu \nu}\epsilon^{abc}=0$.}
\begin{equation}
{\cal L}^{NC}_{YM}=-\frac{1}{2}Tr[W_{\mu \nu}W^{\mu \nu}]-igb^{\alpha \beta}O_{\alpha \beta} \, \
\end{equation}
where
\begin{eqnarray}
{\cal O}^{(6)}_{\alpha \beta}&=& Tr[W_{\alpha \mu}W_{\beta \nu}W^{\mu \nu}] \nonumber \\
&=&\frac{i}{4}\epsilon_{abc}W^a_{\alpha
\mu}W^b_{\beta \nu}W^{c\mu \nu}.
\end{eqnarray}
Here, $W_{\mu \nu}=T^aW^a_{\mu \nu}$ is the gauge tensor
associated with the $SU_L(2)$ group. It is worth to comment that
the symmetric part of this Lorentz 2-tensor, namely ${\cal O}^{(6)
\, S}\equiv g^{\alpha \beta}{\cal O}^{(6)}_{\alpha \beta}$,
induces the anomalous $WW\gamma$ vertex that was considered in
references~\cite{Chia,Peterson,Rizzo,McKellar,Toscano} in the
study of the $b\to s\gamma$ decay. Its phenomenological
implications in other contexts have been studied by diverse
authors~\cite{DA,NT}. In this work, we will study the contribution
of the antisymmetric part of this Lorentz 2-tensor, namely ${\cal
O}^{(6) \, A}\equiv b^{\alpha \beta}{\cal O}^{(6)}_{\alpha
\beta}$, to the $b\to s\gamma$ process. This is the difference
between the study given by the authors of
references~\cite{Chia,Peterson,Rizzo,McKellar,Toscano} and the one
that is presented in this work.
 From the above expressions, it is
easy to derive the anomalous component of the $WW\gamma$ coupling,
which can be written as
\begin{equation}
\mathcal{L}^{NC}_{WW\gamma}= \frac{ie}{2}b^{\alpha
\beta}(W^-_{\alpha \lambda}W^+_{\beta \rho}F^{\lambda
\rho}+W^+_{\alpha \lambda}W^{-\lambda \rho}F_{\beta
\rho}+W^-_{\beta \rho}W^{+\lambda \rho}F_{\alpha \lambda}).
\end{equation}
Using the notation and conventions shown in Fig.\ref{V}, the
corresponding vertex function can be written as follows:
\begin{equation}
\Gamma_{\mu \lambda \rho}(q,k_2,k_3)=\frac{ie}{2}\,b^{\alpha \beta}T^{\eta \xi}_{\mu}\Gamma_{\alpha \beta \eta \xi \lambda \rho}\, \, ,
\end{equation}
where
\begin{equation}
T^{\eta \xi}_{\mu}=q^{\xi}\delta^{\eta}_{\mu}-q^{\eta}\delta^{\xi}_{\mu}
\end{equation}
and
\begin{eqnarray}
\Gamma_{\alpha \beta \eta \xi \lambda \rho}(k_2,k_3) & = & +(k_{2 \beta}g_{\xi \lambda}-k_{2 \xi}g_{\beta \lambda})(k_{3 \alpha}g_{\eta \rho}-k_{3 \eta}g_{\alpha \rho}) {}\nonumber \\
&& {} +g_{\eta \beta}(k_{2 \alpha}g_{\sigma \lambda}-k_{2 \sigma}g_{\alpha \lambda})(k_3^{\sigma}g_{\xi \rho}-k_{3 \xi}\delta^{\sigma}_{\rho}) {}\nonumber \\
&& {} +g_{\eta \alpha}(k_{2 \xi}\delta^{\sigma}_{\lambda}-k_{2 \sigma}g_{\xi \lambda})(k_{3 \beta}g_{\sigma \rho}-k_{3 \sigma}g_{\beta \rho}) \, \, .
\end{eqnarray}
From this expression, it is evident that $\Gamma_{
\lambda \rho \mu}(q,k_2,k_3)$  satisfies the following simple Ward
identities:
\begin{eqnarray}
q^\mu \Gamma_{\lambda \rho \mu}(q,k_2,k_3)=0, \\
k^\lambda_2\Gamma_{\lambda \rho \mu}(q,k_2,k_3)=0, \\
k^\rho_3\Gamma_{\lambda \rho \mu}(q,k_2,k_3)=0.
\end{eqnarray}
The first of this identities guaranties the electromagnetic gauge invariance of the $q_iq_j\gamma$ coupling, whereas the last two are responsible for the gauge independence of this vertex. Electromagnetic gauge invariance means that the photon in the $q_iq_j\gamma$ interaction only can appear through of the $F_{\mu \nu}$ electromagnetic tensor field, which in turns implies that the vertex function associated with this coupling, $\Gamma_{\mu}$, must satisfies the Ward identity $q^\mu \Gamma_\mu=0$. On the other hand, the gauge independence of the $q_iq_j\gamma$ coupling means that the vertex function does not depend on the procedure used to define the propagator of the $W$ gauge boson~\cite{NT}. These facts are still true in the case of an off-shell  $q_iq_j\gamma$ coupling. As we will see below, this transversality of the $\Gamma_{\lambda \rho \mu}$ vertex with respect to the pair of $W$ gauge bosons
is also responsible for the absence of ultraviolet divergences in the $q_i\to q_j\gamma$ decay.

\section{The $q_i\to q_j\gamma$ decay}
\label{D} 
\begin{figure}
\centering\includegraphics[width=2.0in]{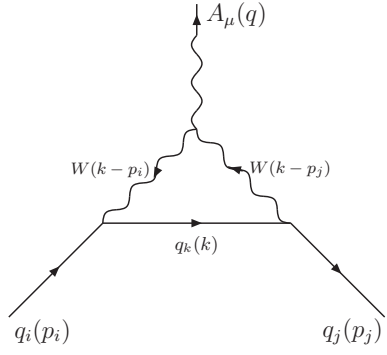}
\caption{\label{FD}Feynman diagram contributing to the flavor violating quark transition $q_i\to q_j\gamma$ in the presence of the background field $b^{\alpha \beta}$.}
\end{figure}

We now turn to calculate the one-loop effect of the
antisymmetric component of ${\cal O}^{(6)}_{\alpha \beta}$ on the
$q_i\to q_j\gamma$ transition. This contribution is given through
the Feynman diagram shown in Fig.~\ref{FD}. Before presenting our
results, let us to argue why one can expect a gauge independent
contribution to the $q_iq_j\gamma$ vertex. In first place is the
fact that the ${\cal O}^{(6)}_{\alpha \beta}$ operator is not
affected by the gauge-fixing procedure of the dimension-four
theory. Also, there are no contributions from Goldstone bosons, as
${\cal O}^{(6)}_{\alpha \beta}$ does not depend on the mechanism
responsible for the electroweak symmetry breaking. Due to this,
all the vertices appearing in the Feynman diagram of Fig.~\ref{FD}
do not depend on the gauge-fixing procedure. In the $R_\xi$-gauge,
explicit gauge dependence only can be carried by the $W$
propagators through their longitudinal components, which however
do not contribute due to the simple Ward identities satisfied by
the $\Gamma_{\lambda \rho \mu}(k_1,k_2,k_3)$ vertex. As we will
see below, this leads to an amplitude that is free of ultraviolet
divergences. After these considerations, we can write the
amplitude for the $q_i\to q_j\gamma $ process as follows
\begin{equation}
{\cal M}=e\overline{u}(p_j,s_j)\Gamma_\mu u(p_i,s_i)\epsilon^{*\mu}(q,\lambda)\, ,
\end{equation}
where $\Gamma_\mu$ is the vertex function characterizing the on-shell $q_iq_j\gamma$ coupling, which can be written as
\begin{equation}
\Gamma_\mu=\frac{g^2}{4}T^{\eta \xi}_\mu b^{\alpha \beta}\sum_{k}V_{ik}V^*_{kj}\int\frac{d^Dk}{(2\pi)^D}\frac{T_{\alpha \beta \sigma \eta \mu }(k,p_i,p_j)}{[k^2-m^2_k][(k-p_i)^2-m^2_W][(k-p_j)^2-m^2_W]}\, ,
\end{equation}
with
\begin{equation}
T_{\alpha \beta \sigma \eta \mu }(k,p_i,p_j)=P_R\gamma^\lambda
\pFMSlash{k} \gamma^\rho \Gamma_{\alpha \beta \eta \xi \lambda \rho
\mu}(k,p_i,p_j)\, .
\end{equation}
Here $\Gamma_{\alpha \beta \eta \xi \lambda \rho
\mu}(k,p_i,p_j)=\Gamma_{\alpha \beta \eta \xi \lambda \rho \mu}(k_2=-k+p_i,k_3=k-p_j)$, $V_{ij}$ are CKM matrix elements, and $P_R=(1+\gamma_5)/2$. It should be noticed that electromagnetic gauge invariance is satisfied automatically due to the presence of the $T^{\eta \xi}_\mu$ tensor as global factor in the vertex function. The evaluation of the above integral is a nontrivial task due to the complicated Lorentz structure of the $WW\gamma$ vertex. However, the calculation can be greatly simplified if we make use of the antisymmetry of the background tensor field $b^{\alpha \beta}$ and also of the $T^{\sigma \eta}_\mu$ tensor, which is antisymmetric in the $\sigma \eta$ indices. Due to this, only the antisymmetric part in these pairs of indices of the $T_{\alpha \beta \sigma \eta \mu }(k,p_i,p_j)$ tensor can contribute to the $\Gamma_\mu$ vertex function. Then, to evaluate $\Gamma_\mu$, we have used only the antisymmetric part of $T_{\alpha \beta \sigma \eta \mu }(k,p_i,p_j)$, which is given by
\begin{equation}
T^{as}_{\alpha \beta \sigma \eta \mu }=\frac{1}{2}\left[\frac{1}{2}\left(T_{\alpha \beta \sigma \eta \mu }-T_{\beta \alpha \sigma \eta \mu }\right)-\frac{1}{2}\left(T_{\alpha \beta  \eta \sigma \mu }-T_{ \beta \alpha \eta \sigma \mu }\right)\right]\, .
\end{equation}
Once solved the loop integral and contracting the result with the $T^{\eta \xi}_\mu$ tensor, one finds that the vertex function can be written in terms of ten  Lorentz tensor gauge structures:
\begin{equation}
\Gamma_\mu=\frac{i\alpha}{8\pi s^2_W}\sum^{10}_{a=1}T^a_\mu \, ,
\end{equation}
where $s_W$ stands for the sine of the weak angle. In the above
expression, the $T^a_{ \mu}$ tensors are electromagnetic gauge
structures in the sense that they satisfy the Ward identity $q^\mu
T^a_{\mu}=0$. Associate with these gauge structures there are form
factors that quantify their relative importance. We now proceed to
present and discuss the main features of these gauge structures
and their associated form factors. We first present the gauge
structures of dipolar type, which are the only present in flavor
changing processes mediated by the photon within the context of
conventional quantum field theory. This component of the $q_i\to
q_j \gamma$ transition can be written as follows:
\begin{equation}
T^1_\mu=-ib^{\alpha\beta}(p_{i\alpha}p_{j\beta}-p_{i\beta}p_{j\alpha})\left(\frac{m_j}{m^2_i-m^2_j}F^1_LP_L+ \frac{m_i}{m^2_j-m^2_i}F^1_RP_R\right)\sigma_{\mu \nu} q^\nu \, ,
\end{equation}
where the dimensionless form factors $F^1_{L,R}$ are given by
\begin{eqnarray}
F^1_L&=&\sum_k V_{ik}V_{kj}^*\left\{-\frac{1}{2}-m^2_WC_0+\frac{1}{2}\frac{m^2_W-m^2_k}{m^2_j}\left[ B_0(2)-B_0(3)\right]-\frac{1}{2}\frac{m^2_W-m^2_k+m^2_i}{m^2_i-m^2_j}\left[B_0(3)-B_0(4) \right] \right\}\, ,\\
F^1_R&=&F^1_L(i\leftrightarrow j)\, .
\end{eqnarray}
 In the above expression, we have introduced the short-hand notation $C_0$ and $B_0(i)$ for the Passarino-Veltman scalar functions of two and three points that characterize the loop effects. The complete amplitude involves the following functions:
\begin{eqnarray}
B_0(1)&=&B_0(0,m^2_W,m^2_W), \\
B_0(2)&=&B_0(0,m^2_k,m^2_W), \\
B_0(3)&=&B_0(m^2_i,m^2_k,m^2_W), \\
B_0(4)&=&B_0(m^2_j,m^2_k,m^2_W),
\end{eqnarray}
\begin{equation}
C_0=C_0(m^2_i,m^2_j,0,m^2_W,m^2_k,m^2_W)\, .
\end{equation}
It is important to notice that these form factors are free of
ultraviolet divergences without invoking the
unitary condition for the CKM matrix $\sum_k
V_{ik}V_{kj}^*=\delta_{ij}$. Of course, those parts of the form
factors which are independent on the internal quark mass $m_k$ do
not contribute to the amplitude, but they have been maintained by
completeness.

There is another gauge structure of dipolar type given by
\begin{equation}
T^2_\mu=-iF^2_L \, b^{\alpha\beta}(\gamma_\alpha q_\beta-\gamma_\beta q_\alpha)P_L\sigma_{\mu\nu}q^\nu \, ,
\end{equation}
where
\begin{equation}
F^2_L=\sum_k V_{ik}V_{kj}^*\left\{\frac{1}{2}m^2_WC_0+\frac{1}{2}\frac{m^2_W-m^2_k}{m^2_i-m^2_j}\left[B_0(3)-B_0(4) \right] +\frac{1}{2}\left[
\frac{m^2_i}{m^2_i-m^2_j}B_0(3)-\frac{m^2_j}{m^2_i-m^2_j}B_0(4) \right]\right\}\, .
\end{equation}
Notice that this form factor is free of ultraviolet divergences only after using the unitarity of the CKM matrix. Also, it is symmetric under the interchange $i \leftrightarrow j$, as must be.

As already mentioned, in conventional quantum field theory the $q_i\to q_j\gamma$ transition only can occurs through a gauge structure of dipolar type. However, when the process occurs in presence of the $b^{\alpha \beta}$ background field, there is a significant increase in the number of gauge structures. We now proceed to list these gauge structures together with their associated form factors:
\begin{equation}
T^3=b^{\alpha
\beta}\left(\frac{m_j}{m^2_i-m^2_j}F^3_LP_L+\frac{m_i}{m^2_j-m^2_i}F^3_RP_R\right)\left(\gamma_\alpha
q_\beta-\gamma_\beta q_\alpha \right)\left( p_i\cdot
q\gamma_\mu-p_{i\mu}\pFMSlash{q}\right)\, ,
\end{equation}
where
\begin{eqnarray}
F^3_L&=&\sum_k V_{ik}V_{kj}^* \frac{1}{4}\left\{1+2m^2_WC_0-\frac{m^2_W-m^2_k}{m^2_j}\left[B_0(2)-B_0(4) \right] +\frac{m^2_i+m^2_W-m^2_k}{m^2_i-m^2_j}\left[B_0(3)-B_0(4) \right]\right\}\, ,\\
F^3_R&=&F^3_L(i\leftrightarrow j)\, ,
\end{eqnarray}

\begin{equation}
T^4_\mu=b^{\alpha\beta}\left(F^4_LP_L+F^4_RP_R \right)\left[ \gamma_\alpha(p_i\cdot q \, g_{\beta\mu}-q_\beta p_{i\mu})-\gamma_\beta(p_i\cdot q \, g_{\alpha\mu}-q_\alpha p_{i\mu})\right]\,
\end{equation}
where
\begin{eqnarray}
F^4_L&=&\sum_k V_{ik}V_{kj}^*\frac{1}{4}\Bigg\{\frac{m_im_j}{m^2_i-m^2_j}\left[B_0(3)-B_0(4) \right] \nonumber \\
&&+\frac{m^2_W-m^2_k}{m_im_j}\left(\frac{m^2_i}{m^2_i-m^2_j}\left[ B_0(2)-B_0(4)\right]-\frac{m^2_j}{m^2_i-m^2_j}\left[ B_0(2)-B_0(3)\right] \right) \Bigg \} \,
\end{eqnarray}
\begin{eqnarray}
F^4_R&=&\sum_k V_{ik}V_{kj}^*\Bigg\{-\frac{1}{6}-m^2_WC_0+\frac{m^2_j}{m^2_i-m^2_j}B_0(4)+\frac{m^2_i}{m^2_j-m^2_i}B_0(3)+\frac{m^2_W+3m^2_k}{4(m^2_i-m^2_j)}\left[B_0(3)-B_0(4)\right] \nonumber \\
&&+\frac{(m^2_W-m^2_k)^2}{4m^2_im^2_j}\left(\frac{m^2_j}{m^2_i-m^2_j}\left[B_0(2)-B_0(3) \right] +\frac{m^2_i}{m^2_j-m^2_i}\left[B_0(2)-B_0(4) \right]\right)
\Bigg \}\, .
\end{eqnarray}
Notice that in this case $F^4_L$ and $F^4_R$ are separately symmetric under the interchange $i \leftrightarrow j$. The following gauge structure is given by
\begin{eqnarray}
T^5_\mu&=&b^{\alpha\beta}\Bigg\{\left[(p_{i\beta}\,g_{\alpha\mu}-p_{i\alpha}\,g_{\beta\mu})\pFMSlash{q}-(p_{i\beta}q_\alpha-p_{i\alpha}q_\beta)\gamma_\mu \right]\left(F^{5i}_LP_L+F^{5i}_RP_R \right) \nonumber \\
&&+\left[(p_{j\beta}\,g_{\alpha\mu}-p_{j\alpha}\,g_{\beta\mu})\pFMSlash{q}-(p_{j\beta}q_\alpha-p_{j\alpha}q_\beta)\gamma_\mu \right]\left(F^{5j}_LP_L+F^{5j}_RP_R \right) \Bigg\}\, ,
\end{eqnarray}
where the $F^{(5i,5j)}_L$ form factors can be decomposed as follows:
\begin{equation}
F^{(5i,5j)}_L=F^{(5iS,5jS)}_L+F^{(5iNS,5jNS)}_L\, ,
\end{equation}
with $S$ and $NS$ stand for symmetric and nonsymmetric under the interchange $i \leftrightarrow j$. These form factors are given by
\begin{eqnarray}
F^{5iS}_L&=&\sum_k V_{ik}V_{kj}^*\Bigg\{\frac{5}{12}-\frac{1}{2}m^2_WC_0+\frac{1}{4(m^2_i-m^2_j)^2} \Bigg(\left[3m^2_im^2_j+m^2_W(m^2_i+m^2_j)\right]\left[B_0(3)+B_0(4) \right] \nonumber \\
&& +3m^2_k\left[m^2_iB_0(3)+m^2_jB_0(4) \right]+3m^2_W\left[m^2_jB_0(3)+m^2_iB_0(4) \right]\Bigg) \Bigg\} \, ,
\end{eqnarray}
\begin{eqnarray}
F^{5iNS}_L&=&\sum_k V_{ik}V_{kj}^*\Bigg\{\frac{m^2_k-m^2_W}{4(m^2_j-m^2_i)}-\frac{m^2_im^2_j-m^2_k(m^2_i+m^2_j)+(m^2_k-m^2_W)(m^2_k+2m^2_W)}{2(m^2_i-m^2_j)}C_0 \nonumber \\
&&+\frac{m^2_i+m^2_j-2(m^2_k+m^2_W)}{2(m^2_i-m^2_j)}B_0(1)-\frac{(m^2_k-m^2_W)^2}{4m^2_i(m^2_i-m^2_j)}B_0(2)\nonumber \\
&&-\frac{m^2_i(m^2_i+6m^2_j)-\left(2-\frac{m^2_j}{m^2_i}\right)(m^2_k-m^2_W)^2}{4(m^2_i-m^2_j)^2}B_0(3) \nonumber \\
&&+\frac{m^4_j-(3m^2_W+m^2_k)(m^2_i+m^2_j)-4(m^2_k+m^2_W)m^2_j-(m^2_W-m^2_k)^2}{4(m^2_i-m^2_j)^2}B_0(4) \Bigg\}\, ,
\end{eqnarray}
\begin{eqnarray}
F^{5jS}_L&=&\sum_k V_{ik}V_{kj}^*\Bigg\{\frac{7}{12}+\frac{3}{2}m^2_WC_0+\frac{1}{(m^2_i-m^2_j)^2}\left[m^4_iB_0(3)+m^4_jB_0(4) \right] \nonumber \\
&&+\frac{4m^2_k(m^2_i+m^2_j)-m^2_im^2_j}{4(m^2_i-m^2_j)^2}\left[B_0(3)+B_0(4)\right]+\frac{5(m^2_W-m^2_k)}{4(m^2_i-m^2_j)^2}\left[m^2_iB_0(3)+m^2_jB_0(4)\right] \Bigg\}\, ,
\end{eqnarray}
\begin{eqnarray}
F^{5jNS}_L&=&\sum_k V_{ik}V_{kj}^*\Bigg\{\frac{m^2_k-m^2_W}{4(m^2_i-m^2_j)}+\frac{m^2_im^2_j-m^2_k(m^2_i+m^2_j)+(m^2_k-m^2_W)(m^2_k+2m^2_W)}{2(m^2_i-m^2_j)}C_0 \nonumber \\
&&+\frac{m^2_i+m^2_j-2(m^2_k+m^2_W)}{4(m^2_j-m^2_i)}B_0(1)-\frac{(m^2_W-m^2_k)^2}{4m^2_j(m^2_j-m^2_i)}B_0(2) \nonumber \\
&&+\frac{m^4_i-(m^2_k+3m^2_W)(m^2_i+m^2_j)-4(m^2_k+m^2_W)m^2_i-(m^2_W-m^2_k)^2}{4(m^2_i-m^2_j)^2}B_0(3)\nonumber \\
&&-\frac{m^2_j(m^2_j+6m^2_i)-\left(2-\frac{m^2_i}{m^2_j} \right)(m^2_W-m^2_k)^2}{4(m^2_i-m^2_j)^2}B_0(4) \Bigg\}\, .
\end{eqnarray}
Notice that $F^{5iNS}_L \leftrightarrow F^{5jNS}_L$ under $i \leftrightarrow j$. Also, it is important to notice that $F^{5i}_L$ and $F^{5j}_L$ are free of ultraviolet divergences only after using $\sum_k V_{ik}V_{kj}^*=\delta_{ij}$. The right-handed form factors are simpler,
\begin{eqnarray}
F^{5i}_R&=&\sum_k V_{ik}V_{kj}^*\Bigg\{\frac{m_im_j}{4(m^2_j-m^2_i)}\left(1+2m^2_WC_0\right)+\frac{m_im_j(m^2_j+m^2_W-m^2_k)}{4(m^2_i-m^2_j)^2}\left[B_0(4)-B_0(3) \right]\nonumber \\
&&-\frac{m_j(m^2_W-m^2_k)}{4m_i(m^2_j-m^2_i)}\left[B_0(2)-B_0(3) \right] \Bigg\}\, , \\
F^{5j}_R&=&F^{5i}_R(i\leftrightarrow j)\, .
\end{eqnarray}
As it is evident, these form factors are free of ultraviolet divergences.

Other gauge structure is given by
\begin{equation}
T^6_\mu=b^{\alpha\beta}(q_\alpha\,g_{\beta\mu}-q_\beta\, g_{\alpha\mu})(m_jF^6_lP_L+m_iF^6_RP_R)\, ,
\end{equation}
where
\begin{eqnarray}
F^6_L&=&\sum_k V_{ik}V_{kj}^*\Bigg\{\frac{2m^2_i+m^2_j+3(m^2_W-m^2_k)}{6(m^2_j-m^2_i)}+\frac{\left[4m^2_i+m^2_j+5(m^2_W-m^2_j)\right]m^2_W}{m^2_j-m^2_i}C_0\nonumber \\
&&+\frac{4m^2_W}{m^2_i-m^2_j}\left[ B_0(1)-B_0(3)\right]-\frac{(m^2_W-m^2_k)(2m^2_j-m^2_i+m^2_W-m^2_k)}{2m^2_j(m^2_j-m^2_i)}\left[B_0(2)-B_0(4)\right]\nonumber \\
&&-\frac{m^2_im^2_j+(m^2_W-m^2_k)^2+(m^2_i-m^2_j)(9m^2_W-m^2_k)}{2(m^2_i-m^2_j)^2}\left[B_0(3)-B_0(4)\right] -\frac{1}{2}B_0(4) \Bigg\}\, , \\
F^6_R&=&F^6_L(i \leftrightarrow j)\, .
\end{eqnarray}
These form factors are finite only after using the $\sum_k V_{ik}V_{kj}^*=\delta_{ij}$ relation, which is evident from the presence of the $B_0(4)$ scalar function at the end of the expression for $F^6_L$.

The remaining four gauge structures are more complicated. To simplify the notation, it is convenient to introduce the following definitions:
\begin{eqnarray}
T^i_{\alpha\beta\mu}&=&(p_{i\alpha}q_\beta-p_{i\beta}q_\alpha)p_{i\mu}-(p_{i\alpha}\,g_{\beta\mu}-p_{i\beta}\,g_{\alpha\mu})p_i\cdot q \\
T^j_{\alpha\beta\mu}&=&(p_{j\alpha}q_\beta-p_{j\beta}q_\alpha)p_{j\mu}-(p_{j\alpha}\,g_{\beta\mu}-p_{j\beta}\,g_{\alpha\mu})p_j\cdot q \\
T^{ij}_{\alpha\beta\mu}&=&(p_{i\alpha}q_\beta-p_{i\beta}q_\alpha)p_{j\mu}-(p_{i\alpha}\,g_{\beta\mu}-p_{i\beta}\,g_{\alpha\mu})p_j\cdot q \\
T^{ji}_{\alpha\beta\mu}&=&(p_{j\alpha}q_\beta-p_{j\beta}q_\alpha)p_{i\mu}-(p_{j\alpha}\,g_{\beta\mu}-p_{j\beta}\,g_{\alpha\mu})p_i\cdot q\,.
\end{eqnarray}
In terms of these tensor structures, one can write
\begin{equation}
T^7_\mu=b^{\alpha\beta}\left(\frac{m_j}{m^2_j-m^2_i}F^{7}_LP_LT^i_{\alpha\beta\mu}+\frac{m_i}{m^2_i-m^2_j}F^{7}_RP_RT^j_{\alpha\beta\mu}\right)\, ,
\end{equation}
where
\begin{eqnarray}
F^{7}_L&=&\sum_k V_{ik}V_{kj}^*\Bigg\{\frac{1}{6(m^2_j-m^2_i)^2}\left\{(m^2_i+m^2_j)^2+20m^2_im^2_j+6m^4_j+6(m^2_W-m^2_k)(2m^2_i+3m^2_j)\right]\nonumber \\
&&+\frac{1}{(m^2_j-m^2_i)^2}\Big[m^2_im^2_j(m^2_i+2m^2_j)+m^2_j\Big((m^2_W+m^2_k)^2+m^2_Wm^2_j\Big)+(m^2_W-m^2_k)\Big(m^2_i(m^2_i+2m^2_j)\nonumber \\
&&+2m^2_j(m^2_j+2m^2_i)+6m^2_W(m^2_i+2m^2_j)+4(m^2_W-m^2_k)(m^2_i+m^2_j)+3(m^2_W-m^2_k)^2\Big)\nonumber \\
&&+m^2_W\left(m^2_i(m^2_i+12m^2_j)+4m^2_j(m^2_j-m^2_k) \right)\Big]C_0\nonumber \\
&&-\frac{1}{2(m^2_j-m^2_i)^2}\Big[m^2_j(m^2_j+5m^2_i)+(m^2_W-m^2_k)\left(7m^2_j+5m^2_i+6(m^2_W-m^2_k)\right)\nonumber \\
&&+4m^2_W(m^2_i+3m^2_j)\Big]\left[B_0(1)-B_0(3)\right] \nonumber \\
&&-\frac{m^2_W-m^2_k}{2m^2_i(m^2_j-m^2_i)^2}\left[m^2_i(3m^2_i+7m^2_j)+(m^2_W-m^2_k)(9m^2_i+m^2_j)\right]\left[B_0(2)-B_0(4) \right]\nonumber \\
&&-\frac{1}{2m^2_i(m^2_j-m^2_i)^3}\Bigg[m^2_im^2_j\left(m^4_i+m^4_j+4m^2_i(m^2_i+4m^2_j)+16m^2_W(m^2_i+m^2_j)\right)\nonumber \\
&&+(m^2_W-m^2_k)\Bigg(m^2_i\left(4(m^4_i+m^4_j)+6m^2_j(m^2_j+5m^2_i) \right)\nonumber \\
&&+(m^2_W-m^2_k)\left(-m^4_i-m^4_j+12m^2_i(m^2_i+m^2_j) \right) \Bigg)\Bigg]\left[B_0(3)-B_0(4) \right]
\Bigg\}\, \\
F^{7}_R&=&F^{7}_L(i \leftrightarrow j)\, .
\end{eqnarray}

\begin{equation}
T^8_\mu=b^{\alpha\beta}\left(\frac{m_j}{m^2_j-m^2_i}F^{8}_LP_LT^j_{\alpha\beta\mu}+\frac{m_i}{m^2_i-m^2_j}F^{8}_RP_RT^i_{\alpha\beta\mu}\right),
\end{equation}
where
\begin{eqnarray}
F^{8}_L&=&\sum_k V_{ik}V_{kj}^*\Bigg\{\frac{1}{6m^2_j(m^2_j-m^2_i)^2}\Big[m^2_j(19m^4_i+10m^2_im^2_j+m^4_j)+3(m^2_W-m^2_k)(m^4_j-m^4_i+10m^2_im^2_j)\Big]\nonumber \\
&&+\frac{3}{(m^2_j-m^2_i)^2}\Big[m^2_i\Big(m^2_im^2_j+2m^2_W(2m^2_i-m^2_j)\Big)-m^4_i(m^2_k-m^2_W)\nonumber \\
&&-(m^2_W-m^2_k)\Big(6m^2_Wm^2_i+(m^2_W-m^2_k+m^2_j)(m^2_W-m^2_k+2m^2_i) \Big) \Big]C_0\nonumber \\
&&-\frac{1}{2m^2_j(m^2_j-m^2_i)^2}\Bigg[m^2_j\Big(2m^2_i(m^2_i+2m^2_j)+m^2_j(m^2_i-m^2_j) \Big)+2m^2_W(m^4_i-m^4_j+8m^2_im^2_j)\nonumber \\
&&+3(m^2_W-m^2_k)m^2_j\Big(3m^2_i+m^2_j+2(m^2_W-m^2_k) \Big) \Bigg]\left[B_0((1)-B_0(3)\right]\nonumber \\
&&-\frac{m^2_W-m^2_k}{2m^4_j(m^2_j-m^2_i)^2}\Big[m^2_j(2m^4_i+m^4_j+7m^2_im^2_j)\nonumber \\
&&+(m^2_W-m^2_k)\Big(5m^2_j(m^2_j+m^2_i)+2m^2_i(m^2_j-m^2_i) \Big) \Big]\left[ B_0(2)-B_0(4) \right]\nonumber \\
&&+\frac{1}{2m^2_j(m^2_j-m^2_i)^3}\Big[m^2_j\Big(m^2_j(m^4_j-m^4_i)-2m^4_i(m^2_i+m^2_j)-6m^2_im^2_j(m^2_j+2m^2_i)\Big)\nonumber \\
&&+2m^2_W\Big(m^6_i-9m^2_im^2_j(m^2_i+m^2_j)+m^6_j \Big)-(m^2_W-m^2_k)m^2_j\Big(17m^4_i+24m^2_im^2_j+3m^4_j\nonumber \\
&&+2(3m^2_j+8m^2_i)(m^2_W-m^2_k) \Big)  \Big]\left[B_0(3)-B_0(4) \right]
\Bigg\} \, ,\\
F^{8}_R&=&F^{8}(i \leftrightarrow j)\,.
\end{eqnarray}

\begin{equation}
T^9_\mu=b^{\alpha\beta}\left(\frac{m_j}{m^2_j-m^2_i}F^{9}_LP_L+\frac{m_i}{m^2_i-m^2_j}F^{9}_RP_R \right)T^{ij}_{\alpha\beta\mu}\, ,
\end{equation}
where
\begin{eqnarray}
F^{9}_L&=&\sum_k V_{ik}V_{kj}^* \Bigg\{ -\frac{1}{6(m^2_j-m^2_i)^2}\Big[7m^4_i+22m^2_im^2_j+m^4_j+6(m^2_W-m^2_k)(3m^2_i+2m^2_j) \Big]\nonumber \\
&&-\frac{1}{(m^2_j-m^2_i)^2}\Big[m^2_im^2_j(m^2_j+2m^2_i)+m^2_W(5m^4_i+12m^2_im^2_j+m^4_j) +(m^2_W-m^2_k)\Big(2m^4_i+6m^2_im^2_j\nonumber \\
&&+m^4_j+6m^2_W(m^2_j+2m^2_i)+(m^2_W-m^2_k)\Big(5m^2_i+4m^2_j+3(m^2_W-m^2_k) \Big) \Big)\Big]C_0\nonumber \\
&&+\frac{1}{(m^2_j-m^2_i)^2}\Big[m^2_i(m^2_i+5m^2_j) +4m^2_W(m^2_j+3m^2_i)+(m^2_W-m^2_k)\Big(7m^2_i+5m^2_j\nonumber \\
&&+6(m^2_W-m^2_k) \Big)\Big]\left[B_0(1)-B_0(3) \right]\nonumber \\
&&+\frac{m^2_W-m^2_k}{2m^2_j(m^2_j-m^2_i)^2}\Big[m^2_j(3m^2_j+7m^2_i)+(m^2_W-m^2_k)(m^2_i+9m^2_j) \Big]\left[B_0(2)-B_0(4) \right]\nonumber \\
&&+\frac{1}{(m^2_j-m^2_i)^3}\Big[m^2_im^2_j(6m^2_i+5m^2_j)+2m^2_W\Big(m^2_i(m^2_i+3m^2_j)+m^2_j(m^2_j+3m^2_i) \Big)\nonumber \\
&&+(m^2_W-m^2_k)\Big(5m^2_i(m^2_i+m^2_j)+3m^2_j(m^2_j+3m^2_i)+(m^2_W-m^2_k)(7m^2_i+4m^2_j) \Big) \Big]\left[B_0(3)-B_0(4) \right]
\Bigg\}\, ,\\
F^{9}_R&=&F^{9}_L(i \leftrightarrow j)\, .
\end{eqnarray}

\begin{equation}
T^{10}_\mu=b^{\alpha\beta}\left(\frac{m_j}{m^2_j-m^2_i}F^{10}_LP_L+\frac{m_i}{m^2_i-m^2_j}F^{10}_RP_R \right)T^{ji}_{\alpha\beta\mu}\, ,
\end{equation}
where
\begin{eqnarray}
F^{10}_L&=&\sum_k V_{ik}V_{kj}^* \Bigg\{ -\frac{1}{(m^2_j-m^2_i)^2}\Big[5m^2_i(m^2_i+m^2_j)+m^2_j(3m^2_i+2m^2_j)+3(m^2_W-m^2_k)(3m^2_i+2m^2_j) \Big]\nonumber \\
&&-\frac{1}{(m^2_j-m^2_i)^2}\Big[m^2_im^2_j(m^2_j+2m^2_i)+2m^2_W\Big(rm^2_i(m^2_i+m^2_j)+m^2_j(2m^2_i+m^2_j) \Big)\nonumber \\
&&+(m^2_W-m^2_k)\Big(m^4_j+6m^2_W(m^2_j+2m^2_i)+(m^2_W-m^2_k)[ 5m^2_i+4m^2_j+3(m^2_W-m^2_k)]\Big) \nonumber \\
&&-2m^2_i\left(m^2_i(m^2_k-3m^2_W) +m^2_j(3m^2_k-5m^2_W)\right)\Big]C_0\nonumber \\
&&+\frac{1}{2(m^2_j-m^2_i)^2}\Big[m^2_i(m^2_i+5m^2_j)+4m^2_W(3m^2_i+m^2_j)\nonumber \\
&&+(m^2_W-m^2_k)\Big(7m^2_i+5m^2_j+6(m^2_W-m^2_k) \Big) \Big]\left[B_0(1)-B_0(3) \right]\nonumber \\
&&-\frac{m^2_W-m^2_k}{2m^2_j(m^2_j-m^2_i)^2}\Big[(m^2_i+m^2_j)(m^2_i+4m^2_j)+(m^2_W-m^2_k)(m^2_i+9m^2_j) \Big]\left[B_0(2)-B_0(4) \right]\nonumber \\
&&+\frac{1}{2(m^2_j-m^2_i)^3}\Big[m^2_i(m^4_i+m^4_j)+10m^2_im^2_j(m^2_i+m^2_j)+4m^2_W\Big(m^2_i(m^2_i+3m^2_j)+m^2_j(m^2_j+3m^2_i) \Big)\nonumber \\
&&+(m^2_W-m^2_k)\Big(m^4_i+m^4_j+10m^2_i(m^2_i+2m^2_j)+6m^2_j(m^2_i+m^2_j)\nonumber \\
&&+2(m^2_W-m^2_k)(7m^2_i+4m^2_j) \Big) \Big]\left[B_0(3)-B_0(4) \right]
\Bigg\}\, ,\\
F^{10}_R&=&F^{10}_L(i \leftrightarrow j)\,.
\end{eqnarray}

\section{Bounding $\Lambda_{NC}$ from $B\to s\gamma$}
\label{B}As commented in the introduction, the $B\to X_s\gamma$ process is particularly sensitive to new physics effects. It results that the GIM cancelation, which is present in all flavor changing neutral current processes, is less severe in this case due to the large top quark mass. At the leading order (LO) in QCD, the $b- s$ transition is described via an operator product expansion based on the effective Hamiltonian
\begin{equation}
H_{eff}=-\frac{4G_F}{\sqrt{2}}V^*_{ts}V_{tb}\sum_{i=1}^{8}C_i(\mu)O_i(\mu),
\end{equation}
where the Wilson coefficients $C_i(\mu)$ are evolved from the
electroweak scale down to $\mu=m_b$ by the Renormalization Group
Equations. The $O_i$ constitute a set of eight renormalized
dimension-six operators~\cite{Review2}. From these, the $O_{1-6}$
represent interactions among four light quarks, which are not of
interest for our purposes. The remainder operators $O_7$ and $O_8$
parametrize the electromagnetic dipolar transition and the
analogous strong dipolar transition, whose contributions to the
$b\to s\gamma$ and $b\to sg$ transitions are dominated by one-loop
effects of the $t$ quark and the $W$ gauge boson. The
corresponding amplitudes can be written as follows:
\begin{equation}\label{ampsm}
\mathcal{M}_{SM}(b\to s\gamma)=-V_{tb}V^*_{ts}\frac{\alpha^{\frac{3}{2}}}{8\sqrt{\pi}s^2_W m^2_W}C_7(\mu)\bar{s}(p_s)\sigma_{\mu\nu}\epsilon^{*\mu}(q,\lambda)q^\nu(m_s P_L+m_b P_R) b(p_b)\, ,
\end{equation}
\begin{equation}
\mathcal{M}_{SM}(b\to sg)=-V_{tb}V^*_{ts}\frac{\sqrt{\alpha_s}\alpha}{8\sqrt{\pi}s^2_W m^2_W}C_8(\mu)\bar{s}(p_s)\sigma_{\mu\nu}\epsilon^{*\mu}_a(q,\lambda)q^\nu T^a(m_s P_L+m_b P_R)b(p_b),
\end{equation}
where $\epsilon^\mu (q,\lambda)$ and $\epsilon^\mu_a(q,\lambda)$ are the polarization vectors of the photon and gluon, respectively. Here, $T^a$ are the generators of the $SU_C(3)$ group, which are normalized as $Tr(T^aT^b)=\delta^{ab}/2$, and $\alpha_s$ is the strong coupling constant.

In our case, the new physics effects associated with the violation of the Lorentz symmetry only contribute to the electromagnetic dipolar transition. According to the results presented in the previous section, the gauge structures $T^1_\mu$ and  $T^2_\mu$ are of dipolar type. As we will show below, the background field tensor  $b^{\alpha\beta}$ couples to the relevant kinematic variables of the process in a simple way. Accordingly, we write the new physics contribution of dipolar type as follows:
\begin{equation}
\mathcal{M}_{NP}(b\to s\gamma)=\frac{\alpha^{\frac{3}{2}}}{2\sqrt{\pi} s^2_W}\bar{s}(p_s)\Bigg[ p_{b\alpha}b^{\alpha\beta}p_{s\beta} \left(\frac{m_s}{m_b^2-m_s^2}F^1_LP_L+\frac{m_b}{m_s^2-m_b^2}F^1_RP_R\right) + \gamma_\alpha b^{\alpha\beta}q_\beta F^2_LP_L \Bigg]\sigma_{\mu\nu}q^\nu\epsilon^{*\mu}(q,\lambda) b(p_b).
\end{equation}
In the context of the effective theory that we are considering, the total theoretical contribution to the $b-s$ transition is given by the sum of the SM contribution and the new physics effects induced by the anomalous $WW\gamma$ vertex:
\begin{equation}
{\cal M}_T={\cal M}_{SM}+{\cal M}_{NP}.
\end{equation}
However, due to the fact that it is reasonable to assume that the
new physics effects is suppressed with respect to the SM
contribution, we will focus on the interference term. From the
Dirac structures of the two types of contributions, one can see
that there is no interference between the SM amplitude and the new
physics one characterized by the $T^2_\mu$ gauge structure. So it
will be ignored in the following. Our main objective in this
section is to get a bound for the $\Lambda_{NC}$ scale. We will
follow closely the analysis given in Ref.~\cite{UP}. The
discrepancy between the theoretical prediction within the SM and
the experimental measurement can be quantified via the following
ratio:
\begin{equation}\label{disc}
R_{EXP-SM}\equiv \frac{\Gamma_{EXP}-\Gamma_{SM}}{\Gamma_{SM}}=\frac{Br_{EXP}(B\to X_s\gamma)}{Br_{SM}(B\to X_s\gamma)}-1,
\end{equation}
 where $\Gamma_{EXP}$ is the experimental decay width of the $B\to X_s\gamma$ transition and  $\Gamma_{SM}$ is the corresponding theoretical prediction of the SM. In addition, $Br_{EXP}$ and $Br_{SM}$ are the respective branching ratios. The current experimental value which is given by the Heavy Flavor Averaging Group (HFAG)~\cite{HFAG} along with the BABAR, Belle and CLEO experimental results is $Br(B \to X_s\gamma) = (3.52\pm 0.23\pm 0.09)\times10^{-4}$ for a photon energy $E_\gamma>1.6$ GeV. On the theoretical side, the SM prediction at the next to next leading order is $Br(B\to X_s\gamma)=(3.15\pm 0.23)\times 10^{-4}$ for $E_\gamma >1.6$ GeV~\cite{SMP}. Using these results, it is found that the discrepancy between theory and experiment is given by $R_{EXP-SM}=0_\cdot117\pm 0_\cdot113$~\cite{UP}. To constraint the $\Lambda_{NC}$ scale, we will assume that the total theoretical prediction, \textit{i.e.}, the SM prediction plus the anomalous $WW\gamma$ vertex contribution, coincides with the experimental value. Thus, we define the ratio
\begin{equation}\label{discota}
R_{TOT-SM}\equiv \frac{\Gamma_{SM+NP}-\Gamma_{SM}}{\Gamma_{SM}}=\frac{Br_{SM+NP}}{Br_{SM}}-1,
\end{equation}
which quantifies the theoretical discrepancy between the effective theory prediction (SM plus new physics effects) and the SM prediction. We now demand that $R_{TOT-SM}\approx R_{EXP-SM}$, which allows us to obtain a bound for $\Lambda_{NC}$. Before doing this, some extra considerations are required. Working out at the LO, which is sufficient for our purposes, the SM QCD corrected contribution can be written as follows:
\begin{equation}\label{ampsmc}
\mathcal{M}_{SM}(b\to s\gamma)=-V_{tb}V^*_{ts}\frac{\alpha^{\frac{3}{2}}}{8\sqrt{\pi}s^2_W m^2_W}C_7^{eff}(m_b)\bar{s}(p_s)\sigma_{\mu\nu}\epsilon^{*\mu}(q,\lambda)q^\nu(m_s P_L+m_b P_R) b(p_b)\, ,
\end{equation}
where $C_7^{eff}(m_b)=0.689\,C_7(m_W)+0.087\,C_8(m_W)$ is the effective Wilson coefficient at the $m_b$ scale~\cite{SMA}. Similarly, the corresponding new physics approximated contribution can be written as follows:
\begin{align}\label{ampnpc}
\mathcal{M}_{NP}(b\to s\gamma)=\frac{\alpha^{\frac{3}{2}}}{2\sqrt{\pi} s^2_W}C_{NP}\bar{s}(p_s)b^{\alpha\beta}&\Bigg(p_{b\alpha}p_{s\beta}\frac{m_s}{m_b^2-m_s^2}F^1_LP_L\nonumber\\
&+p_{b\alpha}p_{s\beta}\frac{m_b}{m_s^2-m_b^2}F^1_RP_R \Bigg)\sigma_{\mu\nu}q^\nu\epsilon^{*\mu}(q,\lambda) b(p_b),
\end{align}
where $C_{NP}=0.689+0.087/Q_e$ and $Q_e$ is the electric charge of the electron.

Once squared the total amplitude and used the assumption $R_{TOT-SM}\approx R_{EXP-SM}$, one obtains the constraint
\begin{equation}
p_{b\alpha}b^{\alpha \beta}p_{s\beta}<2.3\times 10^{-6}\, .
\end{equation}
On the other hand, the background field $b^{\alpha\beta}$ affects the kinematics of the process as follows:
\begin{equation}
p_{b\alpha}b^{\alpha \beta}p_{s\beta}=\frac{1}{\Lambda^2_{LV}}\Big(E_b\, \mathbf{e}\cdot\mathbf{p_s}-E_s\, \mathbf{e}\cdot \mathbf{p_b}+\mathbf{b}\cdot (\mathbf{p_b}\times \mathbf{p_s})\Big)\, ,
\end{equation}
where $E_b$ and $E_s$ are the energies of the $b$ and $s$ quarks. In the rest frame of the $b$ quark, the above expression takes the simple form
\begin{equation}
p_{b\alpha}b^{\alpha \beta}p_{s\beta}=\frac{1}{2}\Big(\frac{m_b}{\Lambda_{LV}}\Big)^2\Big(1-\frac{m^2_s}{m^2_b}\Big)\, \textit{e} \, \cos\chi \,
\end{equation}
where $\chi$ is the angle formed by the $\mathbf{p_s}$ and $\mathbf{e}$ vectors, whereas $\textit{e}$ represents the magnitude of the electric-like constant $\mathbf{e}$ fields. It is interesting to notice that the background $b^{\alpha \beta}$ field naturally links the relevant scale of the process $m_b$ with the new physics scale $\Lambda_{LV}$. Assuming that $\textit{e} \, \cos\chi \sim 1$, we obtain
\begin{equation}
 \Lambda_{LV}>1.96\, \, \mathrm{TeV}.
\end{equation}

It is worth comparing our bounds for the Lorentz violation scale with those obtained in the context of the NCSM. The best place to look for signals of space-time noncommutativity experimentally is in strictly forbidden processes, as the $Z\to \gamma \gamma$ decay~\cite{Z2gW}. The experimental limit on this decay has been used to impose the constraint $\Lambda_{NC}>1$ TeV~\cite{Z2gT} on the noncommutativity scale\footnote{Here, we will use the notation $\Lambda_{NC}$ instead of  $\Lambda_{LV}$, as it is the one used in the context of noncommutative theories.}. Quarkonia decay modes into two photons, which also are strictly forbidden in the SM, have been proposed as a possible signature of space-time noncommutativity, allowing to estimate a $\Lambda_{NC}$ in the range $0.5-1$ TeV~\cite{QT}. On the other hand, the bound $\Lambda_{NC}>80$ GeV was derived from astrophysical considerations~\cite{ASB}. More recently, the bound $\Lambda_{NC}>3$ TeV was obtained with primordial nucleosynthesis~\cite{NST}. Stronger bounds for the noncommutativity scale have been derived in other contexts~\cite{SME-NCSM,OC}. See however reference~\cite{XC}.

\section{Conclusions}
\label{C}The presence of preferred directions in space is an
unmistakeable sign of the Lorentz symmetry violation. In this
work, some phenomenological implications of a constant
antisymmetric tensor field $b^{\alpha \beta}$, which can arise
from general relativity with spontaneous symmetry breaking or from
field theories formulated in a noncommutative space time, were
studied. The gauge and Lorentz structure of the coupling of this
background field with a dimension-six  $SU_L(2)$-invariant and
Lorentz 2-tensor ${\cal O}_{\alpha\beta}$ operator were analyzed.
The one loop implications of the trilinear $WW\gamma$ vertex that
arises from this extended $SU_L(2)$ Yang-Mills sector on the
flavor changing electromagnetic quark transition $q_i\to
q_j\gamma$ were studied. Exact analytical expressions for this
process were derived. It was found that the corresponding vertex
function is characterized by ten electromagnetic gauge structures,
which are gauge independent and free of ultraviolet divergences.
The phenomenological implications at low energies were studied by
applying this general result to the $b\to s\gamma$ decay. Current
experimental data on this process were used to derive the bound
$\Lambda_{LV}>1.96$ TeV on the Lorentz violation scale. This bound
is more stringent than those derived in the context of the
noncommutative standard model from some forbidden processes and
similar to the one derived recently in the same context from a
primordial nucleosynthesis.

\acknowledgments{We acknowledge financial support from CONACYT and
SNI (M\' exico).}

\end{document}